\documentclass{svproc}
\usepackage{graphics,pstricks}
\title{Is star complexity a proxy for information based complexity of
  graphs?}
\titlerunning{Star complexity versus information based complexity}
\author{Russell K. Standish\inst{1}}
\institute{High Performance Coders, Sydney, Australia\\
  \email{hpcoder@hpcoders.com.au},
  \texttt{https://www.hpcoders.com.au}
}

\newcommand{\EcoLab}{{\sffamily\slshape
    \mbox{\raisebox{.5ex}{Eco}\hspace{-.4em}{\makebox[.5em]{L}ab}}}}

\renewcommand{\star}{\mathrm{Star}}
\newcommand{\starbar}{\overline{\mathrm{Star}}}

\begin{document}
\mainmatter
\maketitle
\begin{abstract}

  Information-based complexity (IBC) is a well-defined
  complexity measure of any object given a description in a language
  and a classifier that identifies those descriptions with the object.

  Of course, the exact numerical value will vary according to the
  descriptive language and classifier, but under certain universality
  conditions (eg the classifier identifies programs of a universal
  Turning machine that halt and output the same value),
  asymptotically, the complexity measure is independent of the
  classifier up to a constant of O(1). 

  The hypothesis being investigated in this work that any
  practical IBC measure will similarly be
  asymptotically equivalent to any other practical IBC  measure.

  Standish presented an IBC measure for graphs ${\cal C}$ that encoded
  graphs by their links, and identifies graphs as those that are
  automorphic to each other.

  An interesting alternate graph measure is {\em star complexity},
  which is defined as the number of union and intersection operations
  of basic stars that can generate the original graph. Whilst not an
  IBC itself, it can be related to an IBC (called ${\cal C}^*$) that is
  strongly correlated with star complexity.

  In this paper, 10 and 22 vertex graphs are constructed up to a star
  complexity of 8, and the ${\cal C}^*$ compared emprically with ${\cal C}$.
  Finally, an easily computable upper bound of star complexity is
  found to be strongly related to ${\cal C}$.

\end{abstract}

\section{Introduction}

Information-based complexity (IBC) \cite{Standish01a} is a
well-defined complexity measure of any object given a description in a
language and a classifier that identifies those descriptions with the
object. When the descriptions are halting programs for a universal
Turing machine $U$, and the classifier is the output of $U$ when it
halts, the complexity measure is the negative logarithm of the
well-known Solomonoff-Levin distribution, and is closely related to
the notion of Kolmogorov complexity\cite{Li-Vitanyi97}.

Already in \cite{Standish01a} we see a generalisation of
Solomonoff-Levin complexity, where an observer of a system is
considered a classifier, and complexity of the system can be computed
by counting descriptions in some given language that the observer
considers to describe the system. Even though humans might be
considered to be Turing complete (Turing himself based his notion of
computability on what a human computer could do), in practical
circumstances, humans, and indeed any animal, will rely on short-cuts
to come to a conclusion about what any description means. The
risk of ``hanging'' whilst computing a non-halting description is
counter-productive to the business of living, and so the behaviour of
fully evaluating any description as a Turing machine would do will be
eliminated by evolution.

However, we hypothesise that given a particular classifier, that two
different encoding schemes will asymptotically give the same
complexity values for a given object, even if the classifiers aren't
acting as universal machines, provided the encoding schemes are
universal (able to generate descriptions for all entities in the domain
of interest).

In the current study, we have a graph IBC measure ${\cal C}$ developed
by encoding the edgelist, with the classifier being graph
automorphism\cite{Standish05a}. It turns out this measure is a specific
instance of one introduced by Mowshowitz\cite{Mowshowitz68c,Mowshowitz68d}.

One could compare this directly with star complexity, which is defined
as the minimal number of union and intersection operations of the
elementary stars, but to be truly comparable, we have developed an IBC measure
that encodes the star complexity, and the sequence of operations
generating the graph from the set of elementary stars. In what
follows, we introduce briefly ${\cal C}$, then the notion of star
complexity and the IBC derived from it ${\cal C}^*$.

\section{Graph complexity ${\cal C}$}

The information content, or complexity of a binary string $x$ is
related to the proportion of strings that are interpreted as meaning
the same thing according to a given classifier\cite{Standish01a}.

\begin{equation}\label{complexity}
  {\cal C}(x) = \ell-\log_2 \omega(x)
\end{equation}
where $\ell$ is the length of the string, and $\omega$ the size of the
set of string of length $\ell$ having the same meaning.

In the case of graphs, it is convenient to use a simple edge list, and
the classifier identifies graphs up to automorphisms (ie vertex labels
are ignored), which allows $\omega$ to be computed by means of
algorithms for solving the {\em graph automorphism problem}\cite{Standish05a}.

\section{Star complexity and ${\cal C}^*$}

A number of different graph measures involve counting the minimum
number of operations (eg union or intersections of the edgesets)
required to construct a graph from elementary components. In the case
of star complexity, the elementary components are stars, one for each
vertex as the hub, connected to all other vertices\cite{Jukna13}.

Example star complexities, denoted $\star({\cal G})$ for a
graph ${\cal G}$, are 0 for $S_i$, the star graph connecting a vertex $i$
to the other $n-1$ vertices of the graph (no operations required), 1
for a graph having a single edge connecting vertex $i$ to $j$
(${\cal G}=S_i\cap S_j$) and 2 for the empty graph
(${\cal G}=S_i\cap S_j\cap S_k, \exists i\ne j\ne k$) and $n-2$ for
the full graph (commonly denoted $K_n$, after the German word {\em
  Komplett}).\footnote{Since the edge $i,j$ is only found in stars
  $S_i$ and $S_j$, in order to include all edges, we require at least
  $n-1$ stars, which can be unioned to give $\star(K_n)=n-2$}

Because the complexity measure (\ref{complexity}) has the same value
for the complementary graph $\overline{\cal G}$ (where each edge in ${\cal G}$ is removed
from $K_n$), for comparative purposes we tend to take the minimum
value of $\star({\cal G})$ and $\star(\overline{\cal
  G})$.

Also for comparative purposes, we introduce an IBC measure based on
star complexity. We start by encoding the wordsize needed to store the
number of vertices and the
number of stars, then those values, and then the ``recipe'' for generating the graph in a
reverse polish notation. Symbols for this recipe consist of $\cap$,
$\cup$ which combine the top two elements of the stack, and the
numbers $0\ldots n-1$, which push the number onto the stack. In the
recipe, the number of stack pushes must be one more than the number of
operations, which must equal the star complexity.

The total number of bits for one representation is
\begin{equation}
3\log_2\max(\star({\cal G}), n) + \log_2 (n+2) (2\star({\cal
  G})+1).
\end{equation}
The first term refers to the most compact encoding of the number of vertices
and the star complexity in fixed width fields and a prefix consisting
of $\log_2\max(\star({\cal G}), n)-1$ 1s, followed by a 0 to indicate
the width of that field. The second term is storage required for the
recipe. For large $n$ this tends to being proportional to
$\star({\cal G})$. However, we must subtract from this the logarithm
of the number of such encodings that encode the same graph,
$\omega$.

Unfortunately, there is no algorithm for computing this,
other than brute-force walking all encodings an counting the number
that match each graph, unlike the automorphism algorithm used for
computing (\ref{complexity}).

\begin{equation}\label{complexity*}
  {\cal C}^* = 3\max(\star({\cal G}), n) + \log_2 (n+2) (2\star({\cal G})+1) - \log_2\omega
\end{equation}

\section{$\starbar({\cal G})$: an easy to calculate upper bound on $\star({\cal G})$}

Since walking all encodings is limited in practice to small graphs,
in order to extend this study to larger graphs, we introduced an
algorithm for estimating the star complexity of a given
graph. Firstly, we can find the subgraph given by all vertices
connected to vertices of maximum degree (ie $\{(i,j)\in{\cal G}:
\mathrm{deg}(i)=n\}$. This subgraph can be most efficiently covered
by the stars $\{S_i: \mathrm{deg}(i)=n\}$. The remaining edges can be
covered by the pairwise intersection of the stars centred on the two
vertices connected by the edge. So we have:

\begin{equation}
  {\cal G}=\bigcup_{i: \mathrm{deg}(i)=n}S_i \cup \bigcup_{(i,j)\in{\cal G}:
    \mathrm{deg}(i)<n \& \mathrm{deg}(j)<n} S_i\cap S_j
\end{equation}

Obviously this overcounts the star complexity of the fully connected
graph $K_n$ by one.

Also the second term can be collected using associativity of set
operations. To give an example,
\begin{displaymath}
S_1\cap S_3 \cup S_1\cap S_4 \cup S_2\cap S_3 \cup S_2 \cap
S_4 = (S_1\cup S_2)\cap(S_3\cup S_4)
\end{displaymath}
so in this case, instead of a contribution of 7 to the star complexity
caused by these 4 edges, it is a contribution of 3. Of course we do
not know that we have found the minimum number of set operations
required to generate the full, so strictly speaking, the result of
this algorithm is an overestimate of star complexity, which we call
$\starbar({\cal G})$.

\section{Methods}

We use the \EcoLab{}
package\footnote{https://github.com/highperformancecoder/ecolab,
  version 6.0}, which has code implementing
(\ref{complexity}), leveraging the Nauty\cite{McKay81} library for
solving the graph automorphism problem.

For computing $\star({\cal G})$, we start by computing
recipes above for 1 through to a maximum of number of stars
(determined largely by available resources and patience, but clearly
less than $n(n-1)$), and storing the lowest star count for the
canonical vertex numbering (computed by Nauty). At the same time, the
number of recipes for that star count giving rise to the same graph is
also tracked, for computing $\omega$ above for the purposes of
computing ${\cal C}^*$.

Unfortunately, the number of recipes $NR$ at a given star value $s$ rises
exponentially: for $s>n$, $NR \propto n!n^{s-n}2^{s-1}$, or for $s\le n$,
$NR\propto s!2^{s-1}$. In practice $s<9$ is about what was possible given
the author's computing infrastructure.

The code implementing the walk, and $\starbar$ is included in the
\EcoLab{} package, and optimised for GPUs using the SYCL
API\cite{reinders2023data}. The code was run on a Asus NUC 13, with
Intel i5-1340 CPU with integrated Xe graphics GPU. Running the code on
the GPU resulted in a 23 times speedup, compared with the single
threaded CPU result.

The datasets for walking the 10 vertex graphs and the 22 vertex graphs, as
well as the sampling of randomly generated 1000 vertex graphs to for
${\cal C}$ and $\starbar$ are available in the supplementary
material\cite{Standish25b}. Also included in the supplementary material are the
Ravel\footnote{https://ravelation.net} model files used to analyse the
data and produce the graphs in the results section.
  
\section{Results}

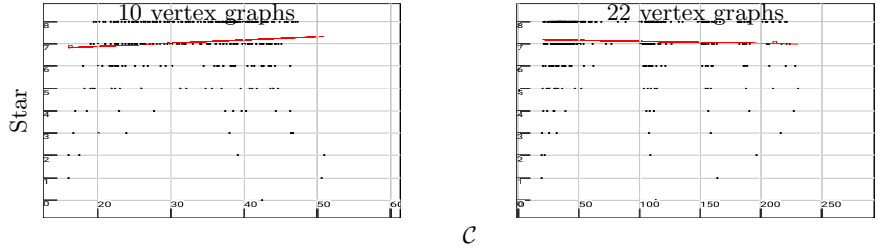
\begin{figure}[h]
  \psset{yunit=3mm}
  \begin{pspicture}(10,10)
    \rput(2.7,5.5){\resizebox{4.75\psxunit}{9.5\psyunit}{\includegraphics{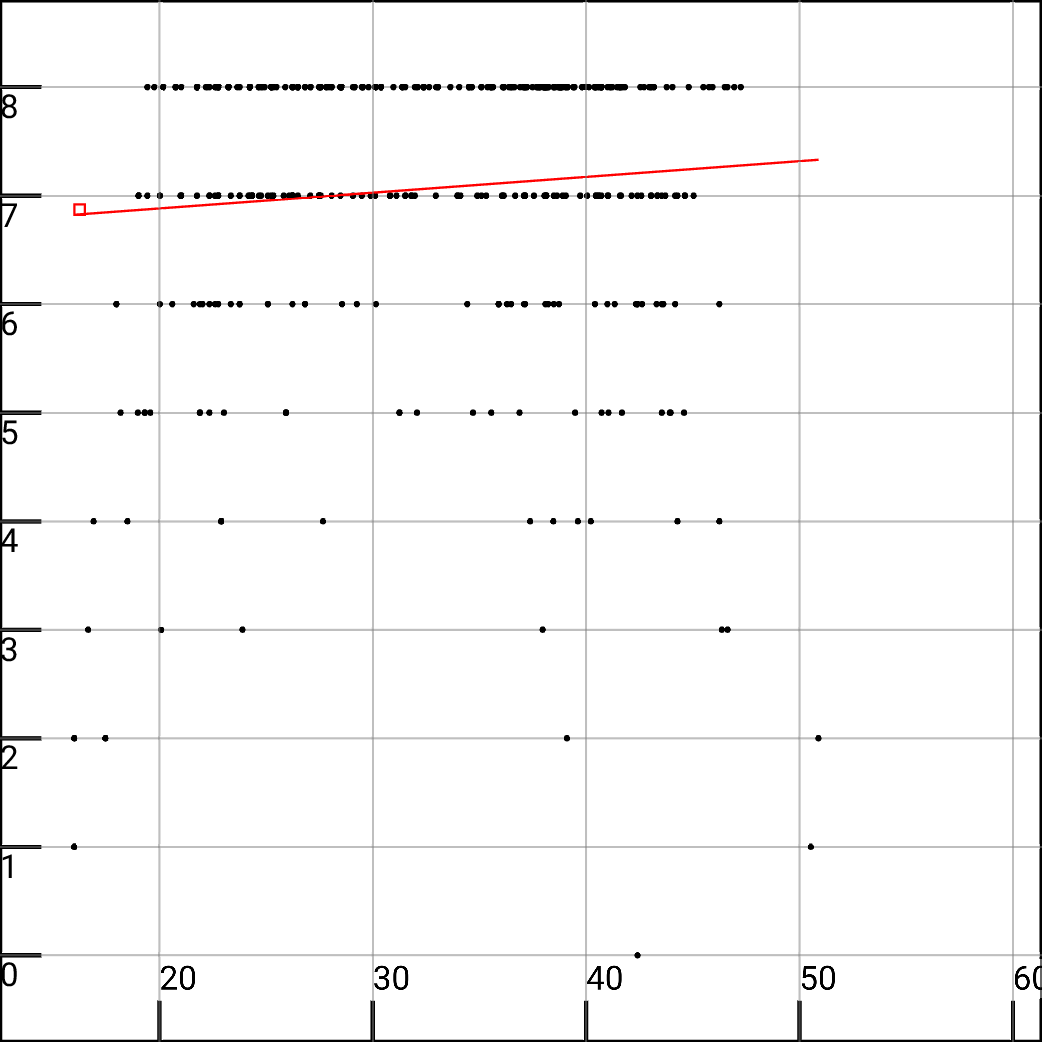}}}
    \rput(9,5.5){\resizebox{4.75\psxunit}{9.5\psyunit}{\includegraphics{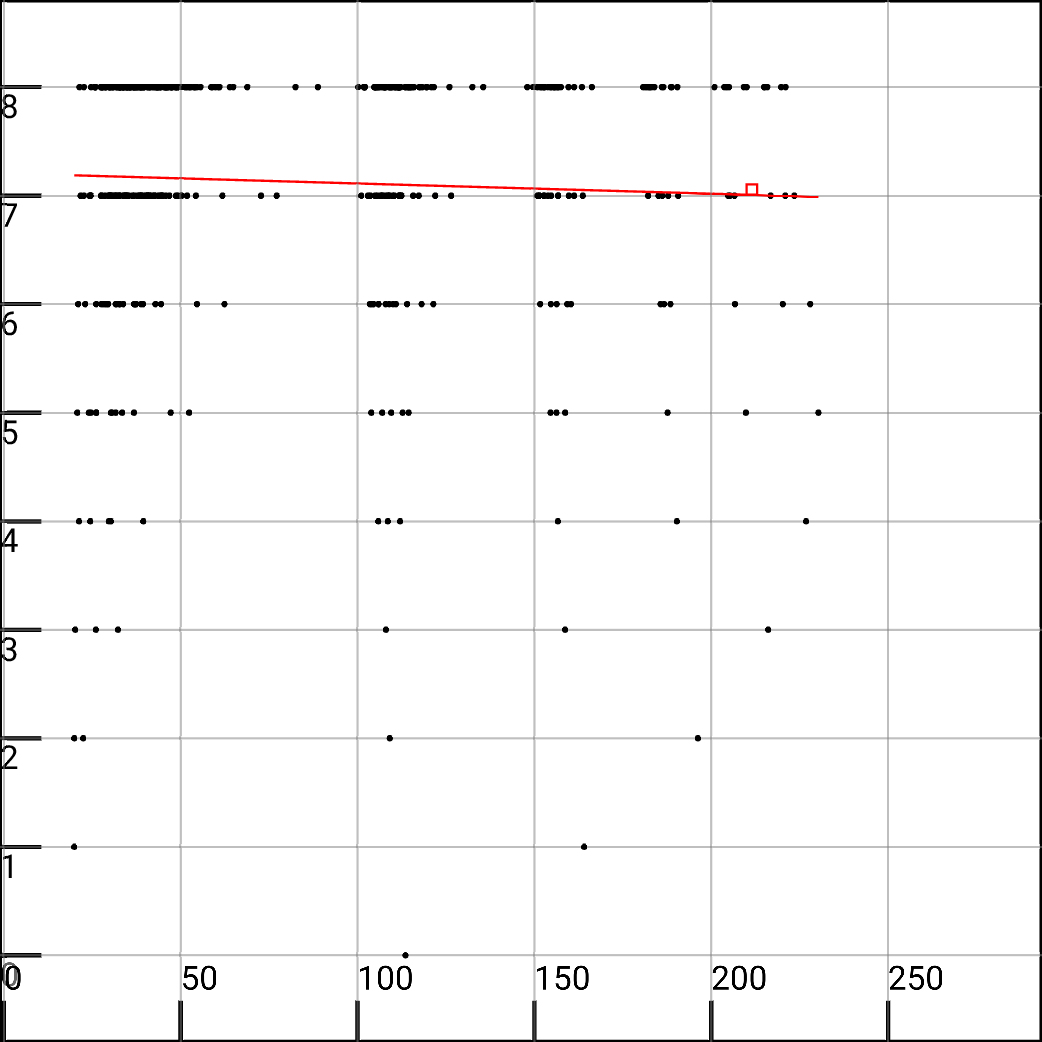}}}
    \rput{90}(0,5.5){Star}
    \rput(6,0){${\cal C}$}
    \rput(2.5,9.8){10 vertex graphs}
    \rput(9,9.8){22 vertex graphs}
  \end{pspicture}
  \caption{Plot of Star vs ${\cal C}$ for (left) all 10 vertex
    graphs and (right) all 22 vertex graphs with star complexity 8
    or less. The line is a linear fit, with correlation coefficient
    $\rho=0.082$ for the 10 vertex graphs, and $-0.041$ for the 22
    vertex graphs.}
  \label{*-C}
\end{figure} 

\begin{figure}[h]
  \psset{yunit=3mm}
  \begin{pspicture}(10,10)
    \rput(2.7,5.5){\resizebox{4.75\psxunit}{9.5\psyunit}{\includegraphics{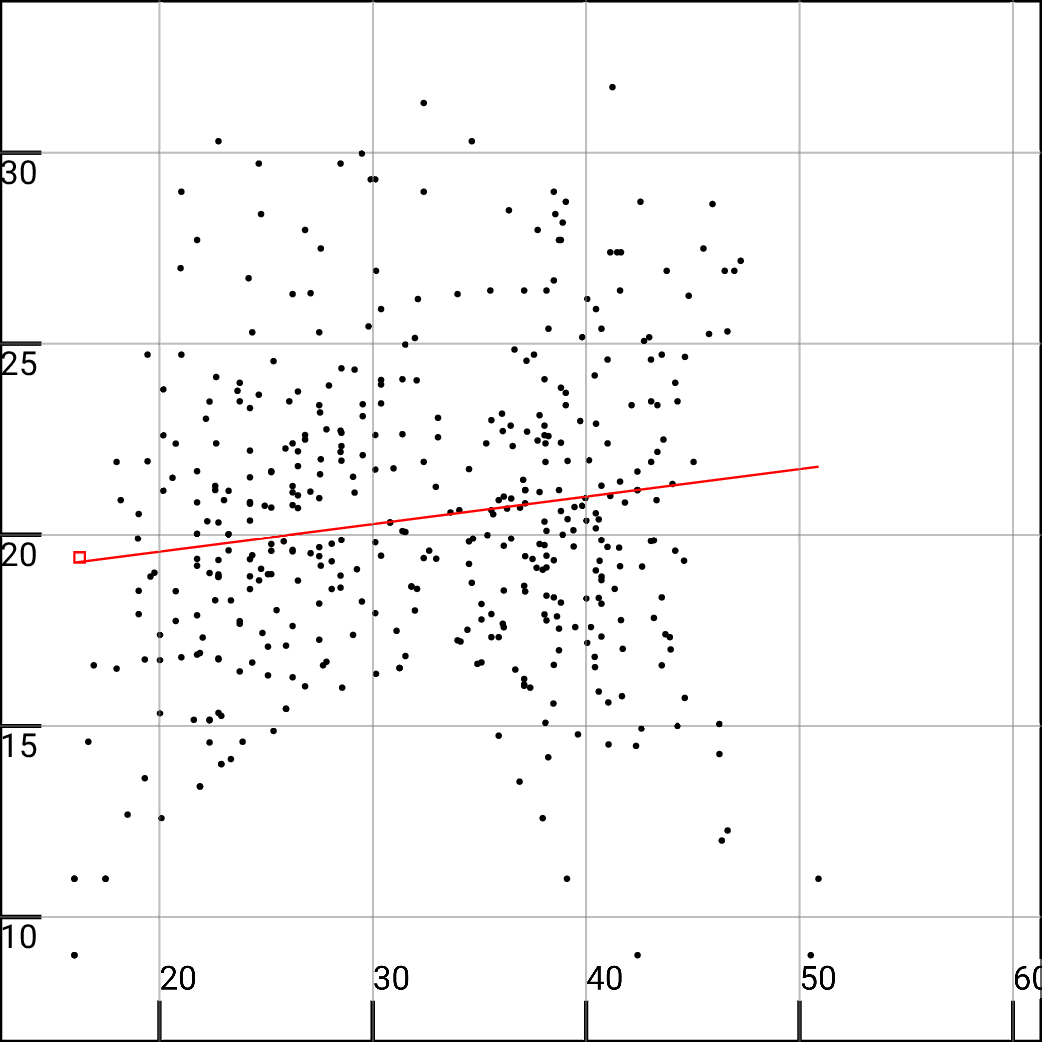}}}
    \rput(9,5.5){\resizebox{4.75\psxunit}{9.5\psyunit}{\includegraphics{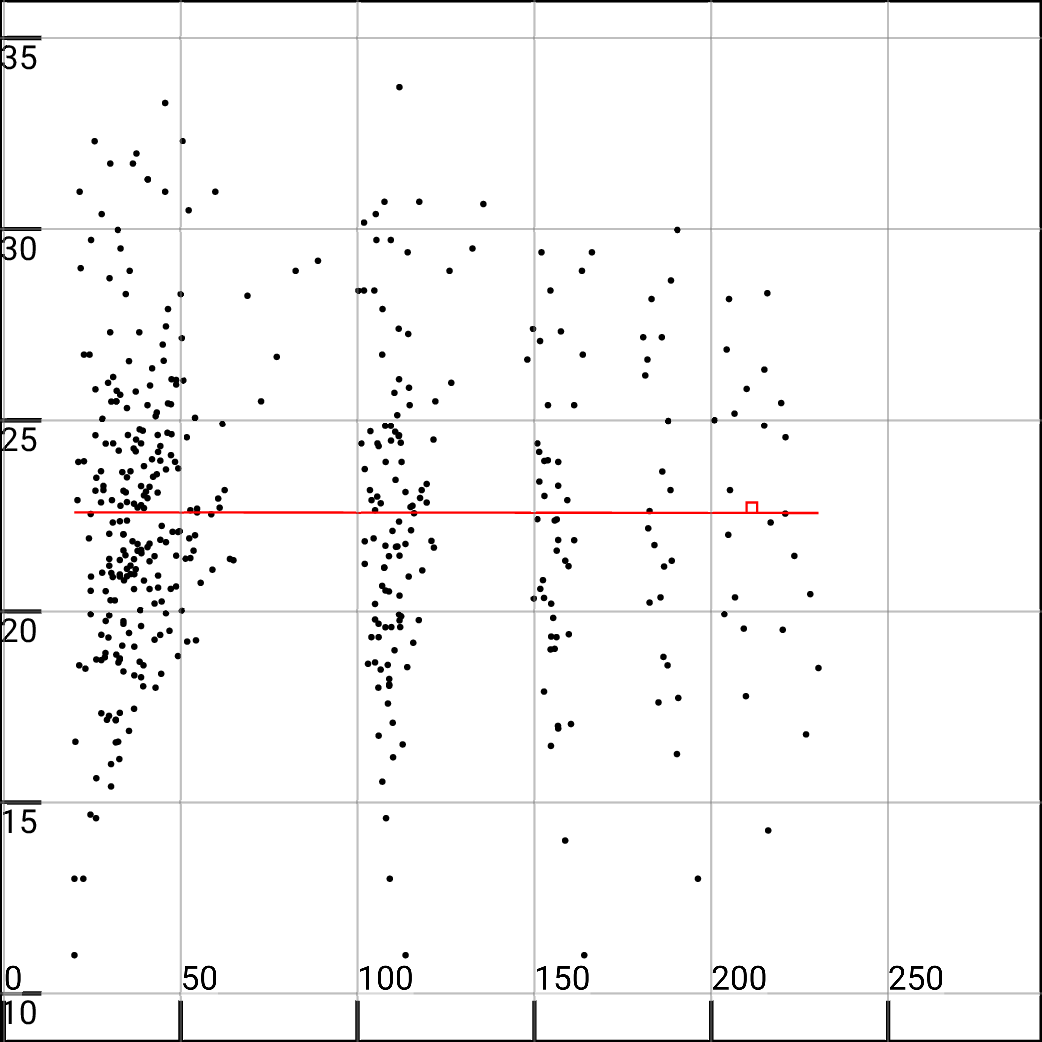}}}
    \rput(0,5.5){${\cal C}^*$}
    \rput(6,0){${\cal C}$}
    \rput(2.5,9.8){10 vertex graphs}
    \rput(9,9.8){22 vertex graphs}
  \end{pspicture}
  \caption{Plot of ${\cal C}^*$ vs ${\cal C}$ for (left) all 10 vertex
    graphs and (right) all 22 vertex graphs with star complexity 8
    or less. The line is a linear fit, with correlation coefficient
    $\rho=0.143$ for the 10 vertex graphs, and $-1.28\times10^{-3}$
    for the 22 vertex graphs}
  \label{C*-C}
\end{figure} 

\begin{figure}[h]
  \psset{yunit=3mm}
  \begin{pspicture}(10,10)
    \rput(2.7,5.5){\resizebox{4.75\psxunit}{9.5\psyunit}{\includegraphics{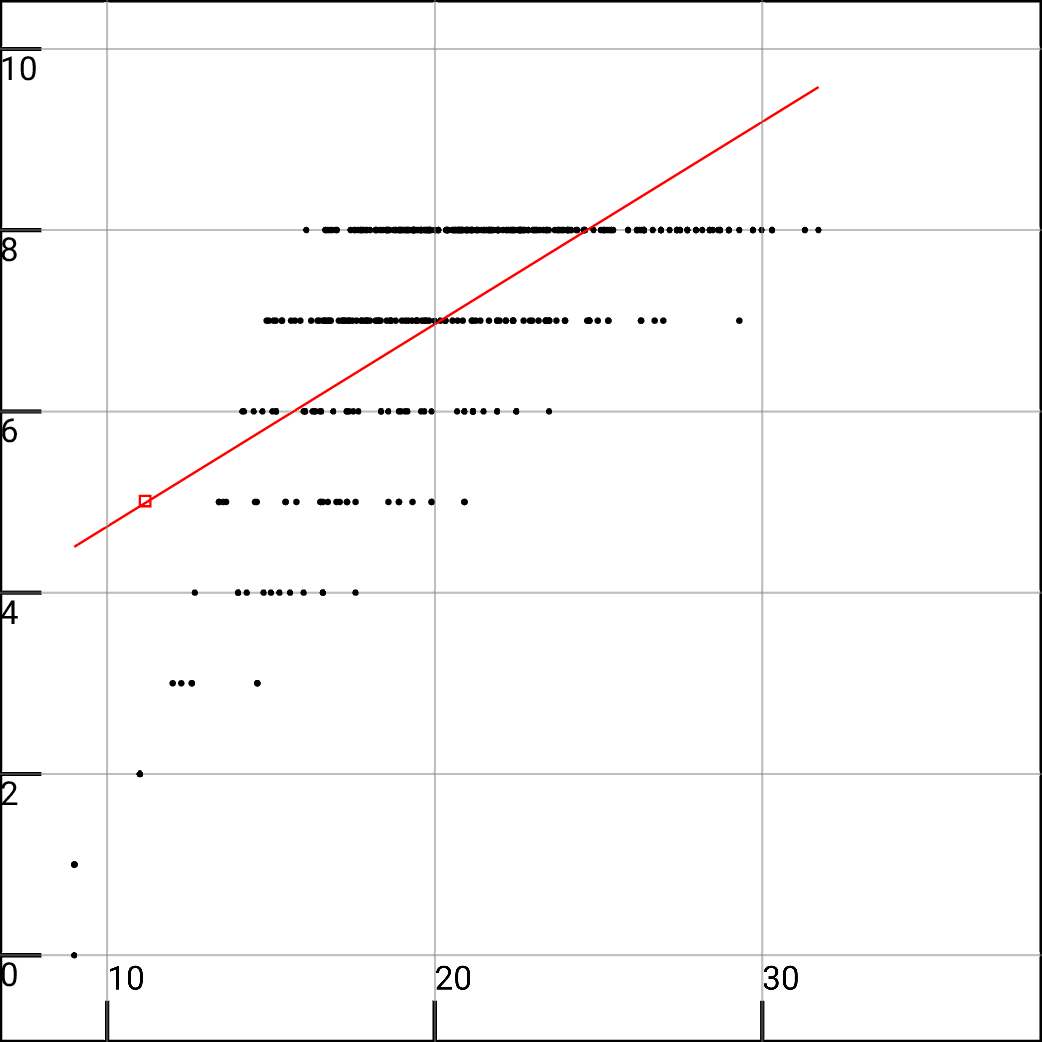}}}
    \rput(9,5.5){\resizebox{4.75\psxunit}{9.5\psyunit}{\includegraphics{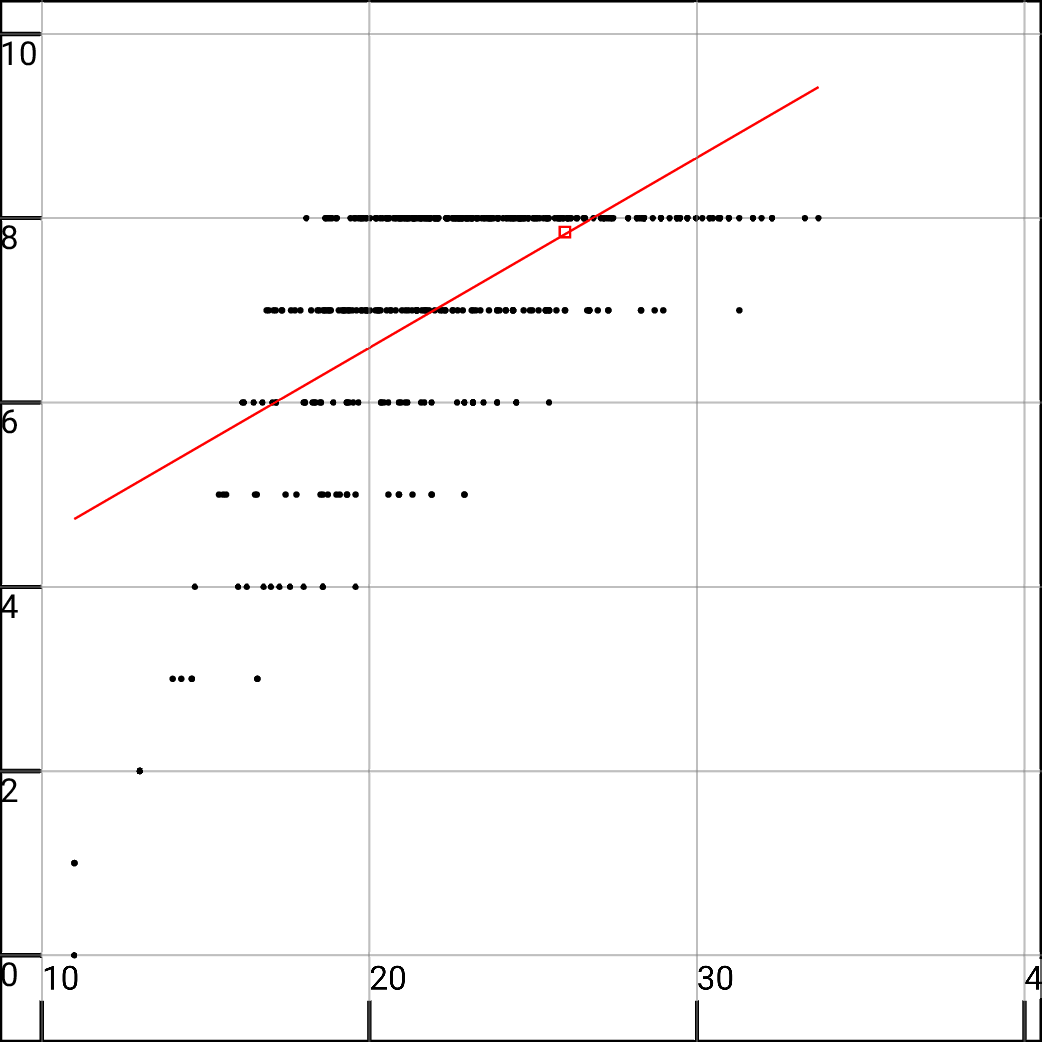}}}
    \rput{90}(0,5.5){Star}
    \rput(6,0){${\cal C}^*$}
    \rput(2.5,9.8){10 vertex graphs}
    \rput(9,9.8){22 vertex graphs}
  \end{pspicture}
  \caption{Plot of $\star$ vs ${\cal C}^*$ for (left) all 10 vertex
    graphs and (right) all 22 vertex graphs with star complexity 8
    or less. The line is a linear fit, with correlation coefficient
    $\rho=0.638$ for the 10 vertex graphs, and $0.611$
    for the 22 vertex graphs}
  \label{*-C*}
\end{figure}
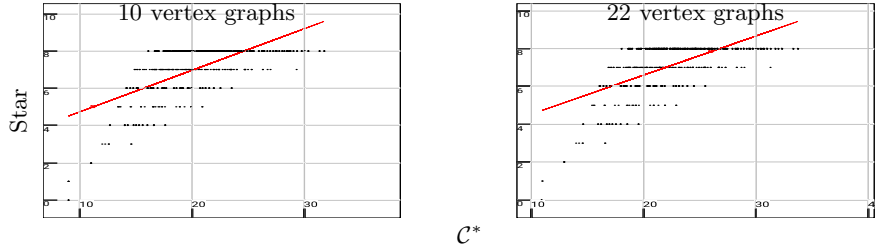

\begin{figure}[h]
  \psset{yunit=3mm}
  \begin{pspicture}(10,10)
    \rput(2.7,5.5){\resizebox{4.75\psxunit}{9.5\psyunit}{\includegraphics{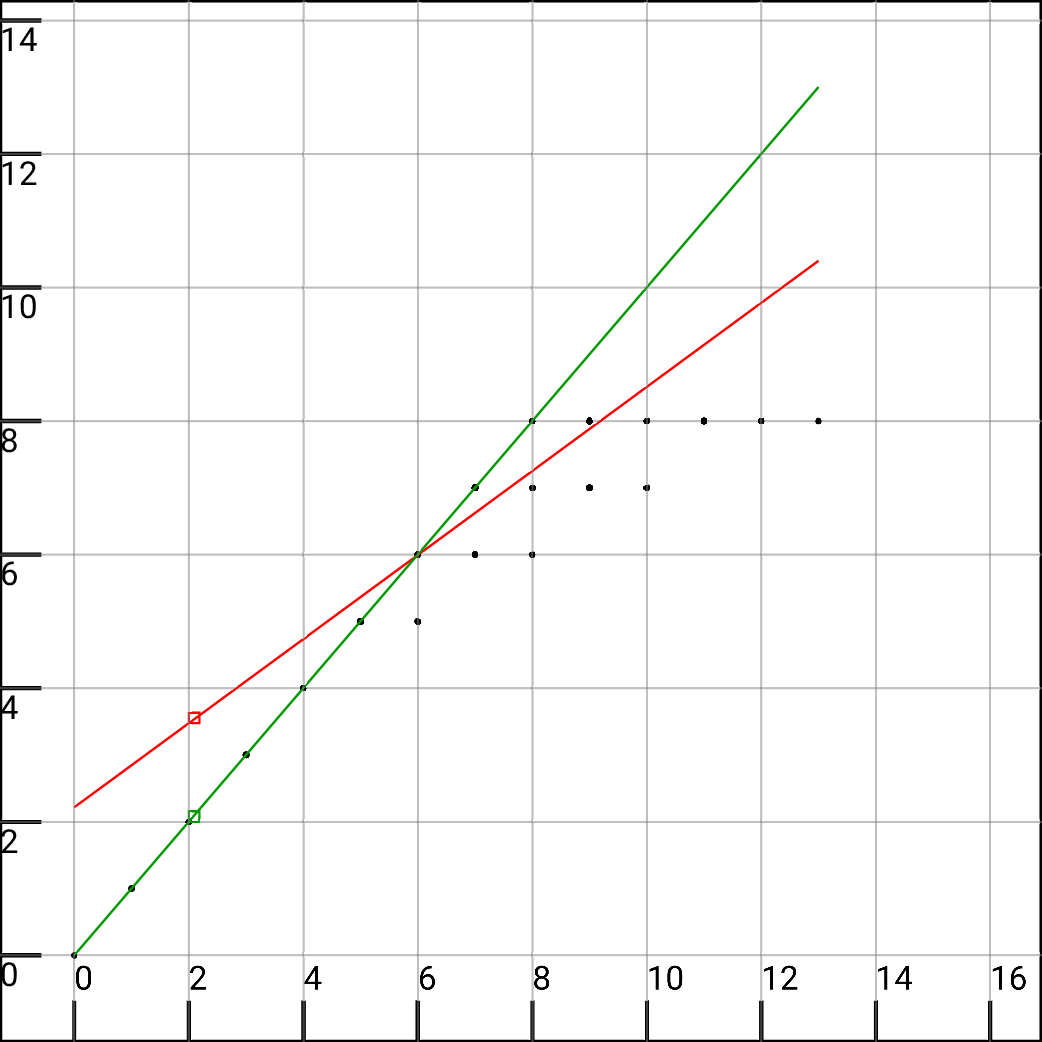}}}
    \rput(9,5.5){\resizebox{4.75\psxunit}{9.5\psyunit}{\includegraphics{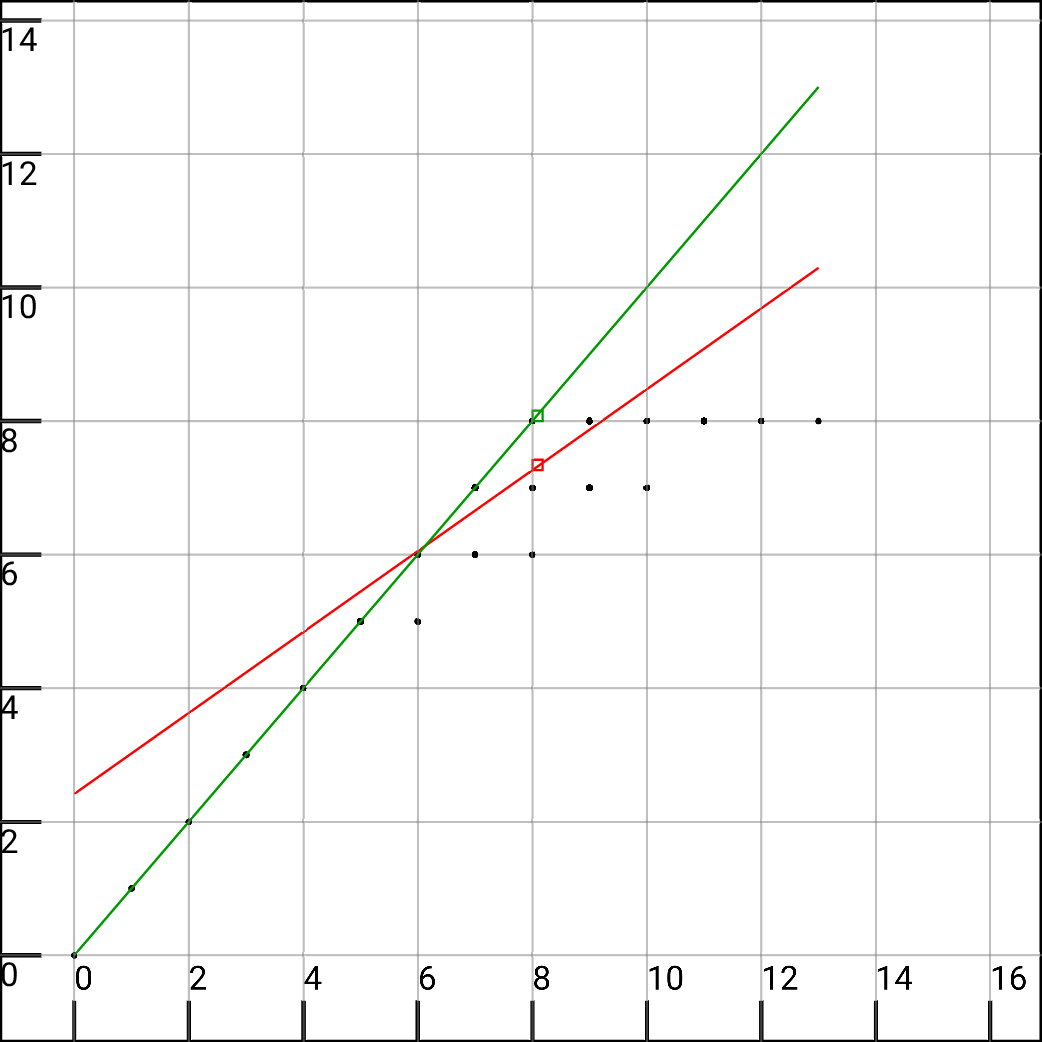}}}
    \rput{90}(0,5.5){Star}
    \rput(6,0){$\starbar$}
    \rput(2.5,9.8){10 vertex graphs}
    \rput(9,9.8){22 vertex graphs}
  \end{pspicture}
  \caption{Plot of $\star$ vs $\starbar$ for (left) all 10 vertex
    graphs and (right) all 22 vertex graphs with star complexity 8
    or less. The red line is a linear fit, with correlation coefficient
    $\rho=0.863$ for the 10 vertex graphs, and $0.850$
    for the 22 vertex graphs. The green line is the identity $\star=\starbar$.}
  \label{star-starbar}
\end{figure}

Figure \ref{*-C} plots the computed star complexity against
${\cal C}$, and figure \ref{C*-C} plots the star complexity derived
information-based complexity measure ${\cal C}^*$ against ${\cal
  C}$. Finally figure \ref{*-C*} plots star complexity against
${\cal C}^*$. ${\cal C}^*$ is quite correlated with star complexity,
as expected, but it does not look promising that star complexity is at
all correlated with complexity. Nevertheless, because of the
difficulty in computing star complexity, only quite small graphs can
be examined, and it may be supposed that star complexity tracks ${\cal
  C}$ when graphs get large enough. To examine that question, we
introduce an easy to compute upper bound $\starbar$. Figure
\ref{star-starbar} plots star complexity against this upper bound, and
it shows the measure is strongly correlated with star complexity ---
indeed we expect that for the vast majority of graphs, the two
measures will be identical. Table \ref{star-starbar-counts} shows the
number of graphs as a function of Star and $\starbar$ for the 22 vertex
case, showing that most graphs lie along the identity diagonal.

\begin{table}
  \begin{tabular}{|rr|rrrrrrrrrrrrrr|}
    \hline
    &&&&&&&&&\rput(0,0){$\starbar$}&&&&&&\\
    &&0&1&2&3&4&5&6&7&8&9&10&11&12&13\\
    \hline
    &0&1&&&&&&&&&&&&&\\
    &1&&2&&&&&&&&&&&&\\
    &2&&&4&&&&&&&&&&&\\
    \rput{-90}(0,0){Star}&3&&&&6&&&&&&&&&&\\
    &4&&&&&11&&&&&&&&&\\
    &5&&&&&&20&3&&&&&&&\\
    &6&&&&&&&36&6&4&&&&&\\
    &7&&&&&&&&70&19&13&6&&&\\
    &8&&&&&&&&&134&49&33&20&7&1\\
    \hline
  \end{tabular}
  \caption{Counts of graphs with particular Star and $\starbar$
    values, out of all 22 vertex graphs with $\star\le 8$. Most graphs
  lie along the $\star=\starbar$ diagonal.}
  \label{star-starbar-counts}
\end{table}

\begin{figure}
  \psset{xunit=12mm,yunit=6mm}
  \begin{pspicture}(10,10)
    \rput(5,5.5){\resizebox{9.5\psxunit}{9.5\psyunit}{\includegraphics{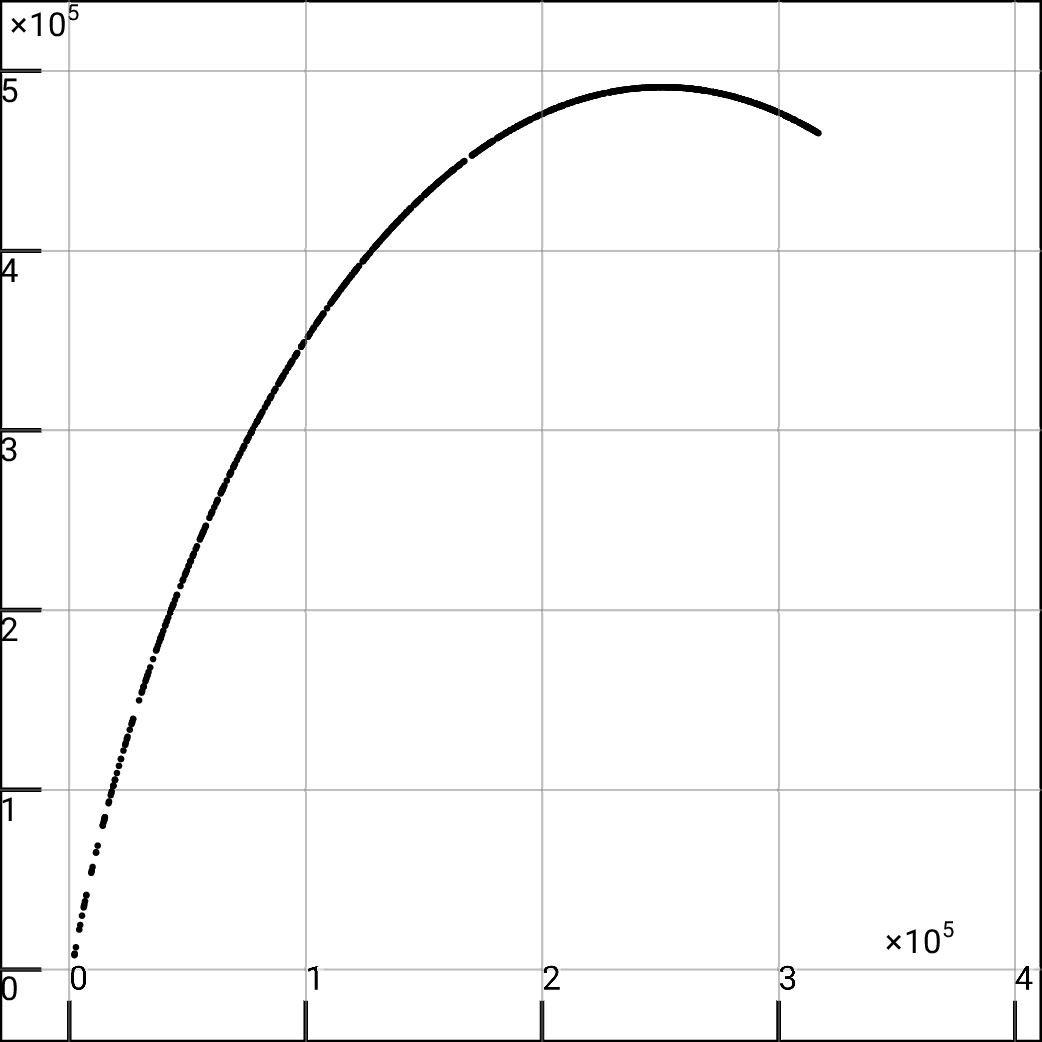}}}
    \rput(0,5.5){${\cal C}$}
    \rput(5,0){$\starbar$}
  \end{pspicture}
  \caption{Plot of ${\cal C}$ against $\starbar$ for 1000 Erd\"os-Renyi
  randomly generated graphs of 1000 vertices each.}
  \label{ER-1000}
\end{figure}

These results give us confidence that the last experiment shown in
figure \ref{ER-1000}, which plots
${\cal C}$ against $\starbar$ for 1000 Erd\"os-Renyi random graphs
with 1000 vertices, demonstrates a strong relationship, albeit a
nonlinear one, between star complexity and graph complexity,
answering the titular question in the affirmative.

\section{Conclusion}

The final experiment of comparing an upper bound of star complexity
called $\starbar$ with the information-based complexity ${\cal C}$
with a large sample of Erd\"os-Renyi generated graphs with 1000
vertices, gave empirical evidence that star complexity and Standish's
information based complexity are indeed measuring the same thing, for
most graphs. One may speculate that other graph complexity measures,
such as linear complexity\cite{Neel-Orrison13} will also be
found to measure the same thing.

\bibliographystyle{spmpsci}
\bibliography{rus}

\end{document}